\documentclass[12pt]{elsart}

\usepackage[latin1]{inputenc}
\usepackage{amsmath}
\usepackage{amsfonts}
\usepackage{amssymb}
\usepackage{ulem}
\usepackage[dvips]{graphics}
\usepackage{epsfig}

\begin{document}

\begin{frontmatter}

\title{Self-synchronization of Cellular Automata:
an attempt to control patterns}
\author[jrs]{J.R. Sánchez},
\ead{jsanchez@fi.mdp.edu.ar}
\author[rlr]{R. L\'{o}pez-Ruiz}
\ead{rilopez@unizar.es}

\address[jrs]{Fac. Ingenier\'{\i}a, Universidad Nacional de Mar del Plata, 
Justo 4302, 7600 Mar del Plata, Argentine,}
\address[rlr]{DIIS and BIFI, Facultad de Ciencias, 
Universidad de Zaragoza, 50009 - Zaragoza, Spain.}

\begin{abstract} 
The searching for the stable patterns in the evolution of 
cellular automata is implemented using stochastic synchronization 
between the present structures of the system
and its precedent configurations. 
For most of the known evolution rules with complex 
behavior a dynamic competition among all the possible stable 
patterns is established and no stationary regime is reached.
For a particular rule coded by the decimal number 18, 
a self-synchronization phenomenon can be obtained,
even when strong modifications to the synchronization method are applied. 

\end{abstract}


\end{frontmatter}

\section{Introduction}

Cellular automata (CA) are extended systems, discrete both in space and time.  
The simplest one dimensional version of a cellular automaton
is formed by a lattice of $N$ sites or cells, 
numbered by an index $i=1,\ldots,N$. In each site a local variable $\sigma_i$   
taking a binary value, either $0$ or $1$, is asigned.  The binary string 
$\sigma(t)$ formed by all sites values at time $t$ 
represents a configuration of the system.  The system evolves in time 
by the application of a rule $\Phi$. A new configuration $\sigma(t+1)$ is obtained
under the action of the rule $\Phi$ on the state $\sigma(t)$.
Then, the evolution of the automata can be writen as
\begin{equation}
\sigma(t+1) = \Phi\:[\sigma(t)].
\end{equation}
If coupling among nearest neighbors is used, 
the value of the site $i$, $\sigma_i(t+1)$, at time $t+1$ is a function 
of the value of the site itself at time $t$, $\sigma_i(t)$, and the values of 
its neighbors $\sigma_{i-1}(t)$ and $\sigma_{i+1}(t)$ at the same time.
Then, the local evolution is expressed as
\begin{equation}
\sigma_i(t+1) = \phi(\sigma_{i-1}(t),\sigma_i(t),\sigma_{i+1}(t)),
\end{equation}
being $\phi$ a particular realization of the rule $\Phi$.
For such particular implementation, there will be $2^3$ different local input 
configurations for each site and, for each one of them, a binary value can be 
assigned as output. Therefore there
will be $2^8$ different rules $\phi$. Each one of these rules produces a 
different dynamical evolution.  In fact, the dynamical behaviors generated 
by the $256$ rules were already classified in four great groups. 
The reader interested in the details of such classification is addressed to the 
original reference where this development can be followed~\cite{wolfram}.  

CA provide us with simple extended systems in which to prove different methods 
of synchronization.  A stochastic synchronization technique was introduced 
by Morelli and Zanette~\cite{zanette} that works in synchronizing 
two CA evolving under the same rule $\Phi$. The two CA are started 
from different initial conditions  
and they are supposed to have partial knowledge about each other.
In particular, the CA configurations, $\sigma^1(t)$ and $\sigma^2(t)$,  
are compared at each time step.
Then, a fraction $p$ of the total different sites are made equal (synchronized).  
The synchronization is stochastic since the location of the sites that are going to 
be equal is decided at random.
Hence, the dynamics of the two coupled CA, $\sigma(t)=(\sigma^1(t), \sigma^2(t))$, 
is driven by the successive application of two operators:  
\begin{enumerate}
	\item the deterministic operator given by the CA evolution rule $\Phi$,
	 $\Phi[\sigma(t)]=(\Phi[\sigma^1(t)], \Phi[\sigma^2(t)])$, and
	\item the stochastic operator $\Gamma_p$ that produces the result $\Gamma_p[\sigma(t)]$, 
	in such way that, if the sites are different ($\sigma_i^1\neq\sigma_i^2$), then $\Gamma_p$ 
	sets both sites equal to $\sigma_i^1$ or $\sigma_i^2$ with the same probability $p/2$.  
	In any other case $\Gamma_p$ leaves the sites unchanged.  
\end{enumerate}
Therefore the temporal evolution of the system can be written as
\begin{equation}
\sigma(t+1) = (\Gamma_p\circ\Phi)[\sigma(t)] = \Gamma_p[ (\Phi[\sigma^1(t)], \Phi[\sigma^2(t)])].
\end{equation}

\begin{figure}
\includegraphics[angle=-90, width=15cm]{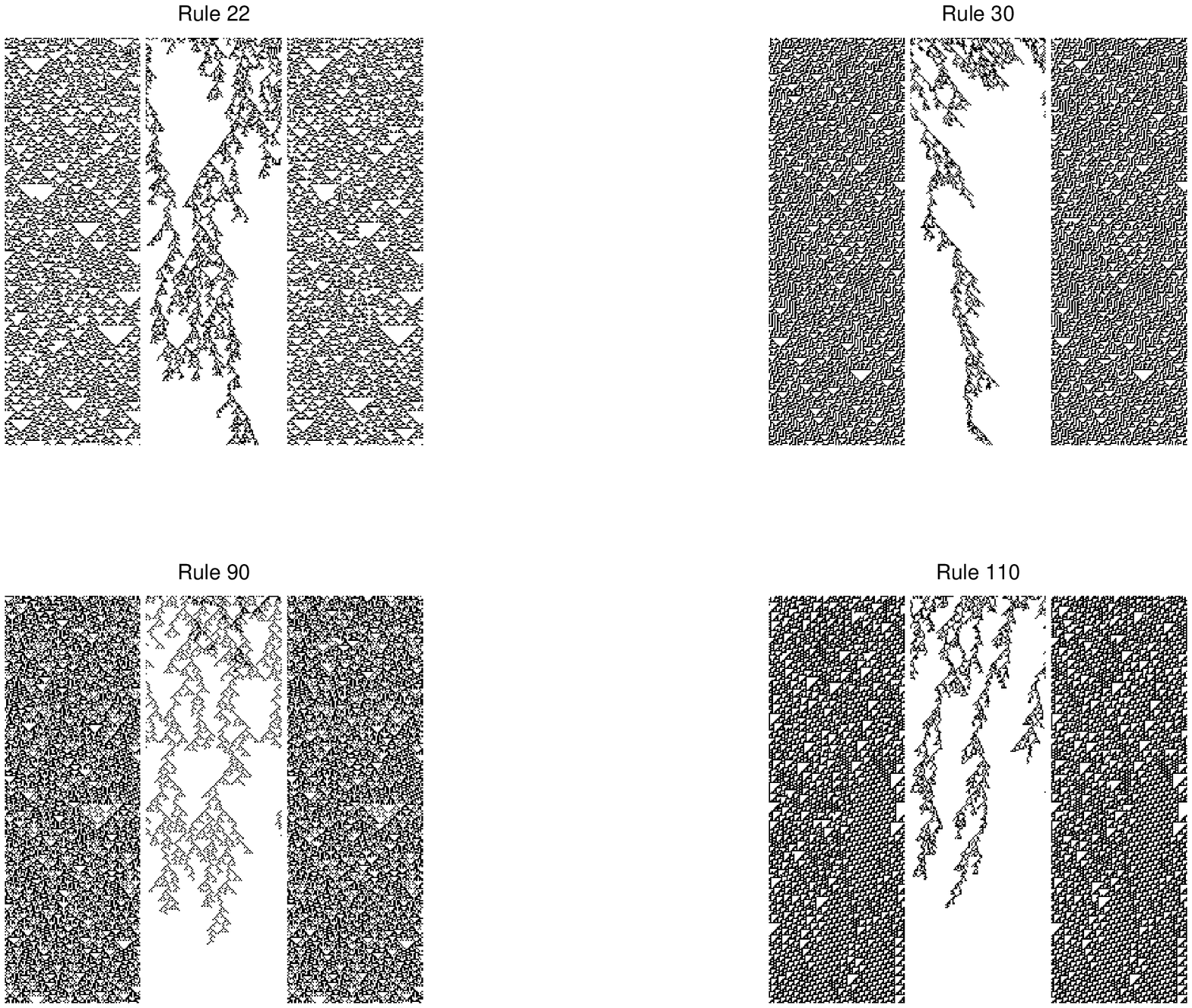}
\caption{Spatio-temporal patterns near the synchronization transition.
Lateral panels represent both CA evolving in time and the central panel displays the 
evolution of the corresponding DA. Time goes from top to bottom.
Lattice size is $N=100$ and the number of iterations is $T=250$. 
The stochastic coupling between both CA is $p=0.23$.}
\label{fig1}
\end{figure}

A simple way to visualize the transition to synchrony can be done by displaying 
the evolution of the difference automaton (DA),
\begin{equation}
\delta_i(t)=\mid\sigma_i^1(t)-\sigma_i^2(t)\mid.
\end{equation}
The mean density of active sites for the DA
\begin{equation}
\rho(t)={1\over N}\sum_{i=1}^{N}\delta_i(t),
\end{equation}
represents the Hamming distance between the automata and 
verifies $0\leq\rho\leq 1$. 
The automata will be synchronized when $\lim_{t\rightarrow\infty}\rho(t)=0$. 
As it has been described in Ref. \cite{zanette},
two different dynamical regimes, controlled by the parameter $p$, can be found 
in the system behavior: 
\begin{eqnarray*}
p<p_c & \rightarrow & \hbox{$\:\lim_{t\rightarrow\infty}\rho(t) \neq 0$ (no synchronization),} \\
p>p_c & \rightarrow & \hbox{$\:\lim_{t\rightarrow\infty}\rho(t)=0$ (synchronization)},
\end{eqnarray*}
being $p_c$ the parameter for which the transition to the synchrony occurs.  
When $p \lesssim p_c$ complex structures can be observed in the DA time evolution.  
In Fig. \ref{fig1} typical cases of such behavior are shown near the synchronization transition. 
Lateral panels represent both CA evolving in time and the central panel displays the 
evolution of the corresponding DA.  When $p$ comes close to the critical value $p_c$ the evolution of
$\delta(t)$ becomes more rarefied and reminds the problem of structures trying to percolate
in the plane~\cite{pomeau86}. A method to detect this kind of transition, based
in the calculation of a statistical measure of complexity for patterns, 
has been proposed in the Refs.~\cite{sanchez1}.

\section{Self-synchronization of cellular automata}

\subsection{First self-synchronization method}

Let us now take a single cellular automaton \cite{margolus}. If $\sigma^1(t)$ is the state of the automaton
at time t, $\sigma^1(t)=\sigma(t)$, and $\sigma^2(t)$ is the state obtained from the application 
of the rule $\Phi$ on that state, $\sigma^2(t)=\Phi[\sigma(t)]$, then the operator $\Gamma_p$ can be 
applied on the pair $(\sigma^1(t), \sigma^2(t))$, giving rise to the evolution law
\begin{equation}
	\sigma(t+1) = \Gamma_p[(\sigma^1(t), \sigma^2(t))] = \Gamma_p[ (\sigma(t), \Phi[\sigma(t)]) ].
	\label{eq-auto}
\end{equation}
The update of the different sites between $\sigma^1(t)$ and $\sigma^2(t)$ takes place with 
probability $p$ in $\sigma^2(t)$. This updated array is the one that stays as $\sigma(t+1)$. 

\begin{figure}
\includegraphics[angle=0, width=15cm]{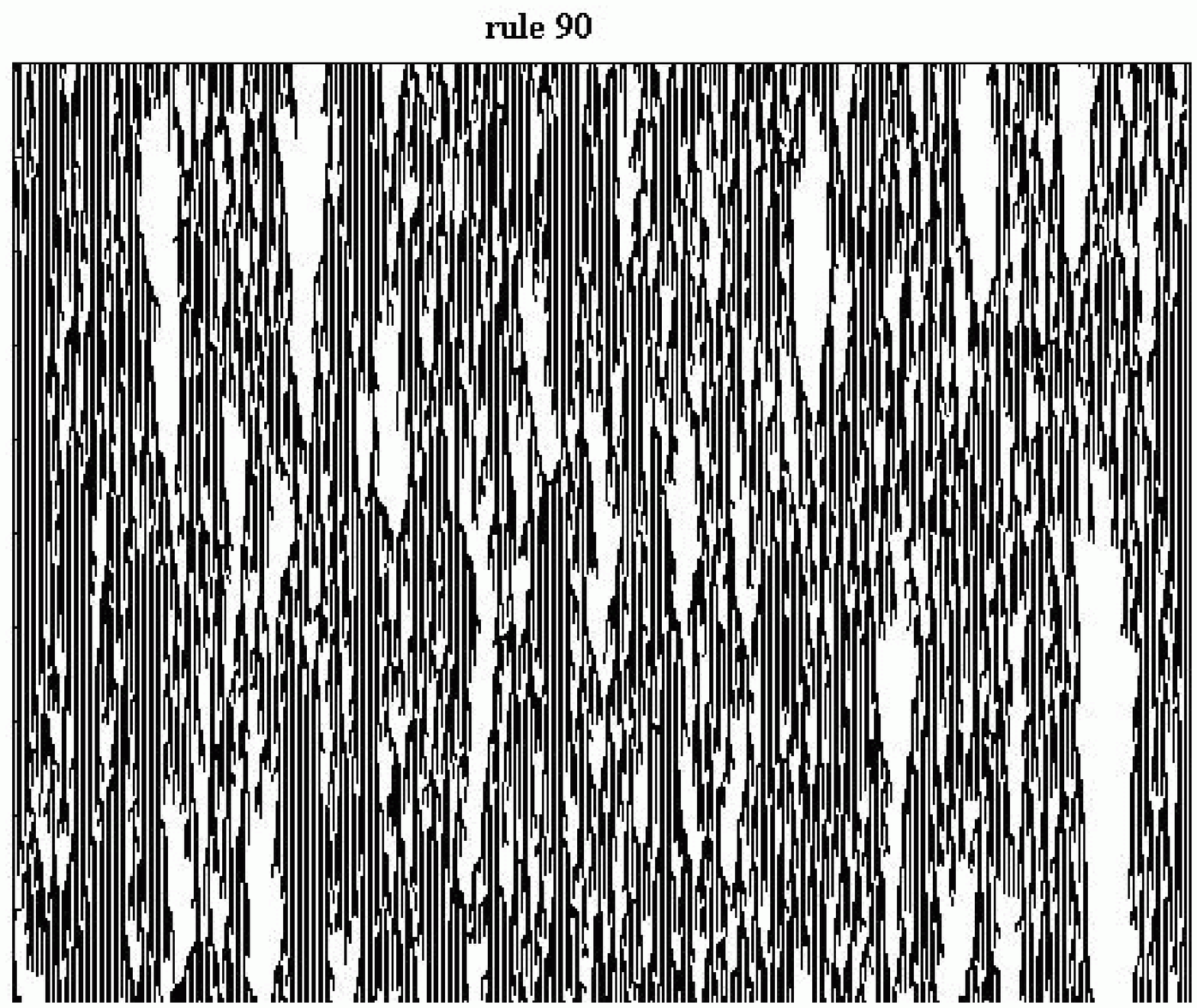}
\caption{Rule $90$ has two stable patterns: one repeats the $011$ string
and the other one the $00$ string. Such patterns are reached by the first 
self-synchronization method but there is a dynamical competition between them.
In this case $p=0{.}9$.
Binary value $0$ is represented in white and $1$ in black.
Time goes from top to bottom.}
\label{fig5}
\end{figure}

It is worth to observe that if the system is initialized with
a stable configuration for the rule $\Phi$, $\Phi[\sigma]=\sigma$, then this state $\sigma$ is also 
stable for the dynamic equation (\ref{eq-auto}), and the evolution will produce
a pattern constant in time. In particular, such behavior can be observed in rule $18$
where the system evolves toward its only stable null pattern.     
However, in general, this stability is marginal. A small modification of the 
initial condition gives rise to patterns variable in time. In fact, as the parameter $p$ increases, 
a competition among the different marginally stable structures takes place.  
The dynamics makes the system to stay close to those states,
although oscillating continuously and randomly among them.
Hence, a complex spatio-temporal behavior is obtained.
Some of these patterns can be seen in Fig. \ref{fig5}.

\subsection{A second self-synchronization method}

\begin{figure}
\includegraphics[angle=0, width=15cm]{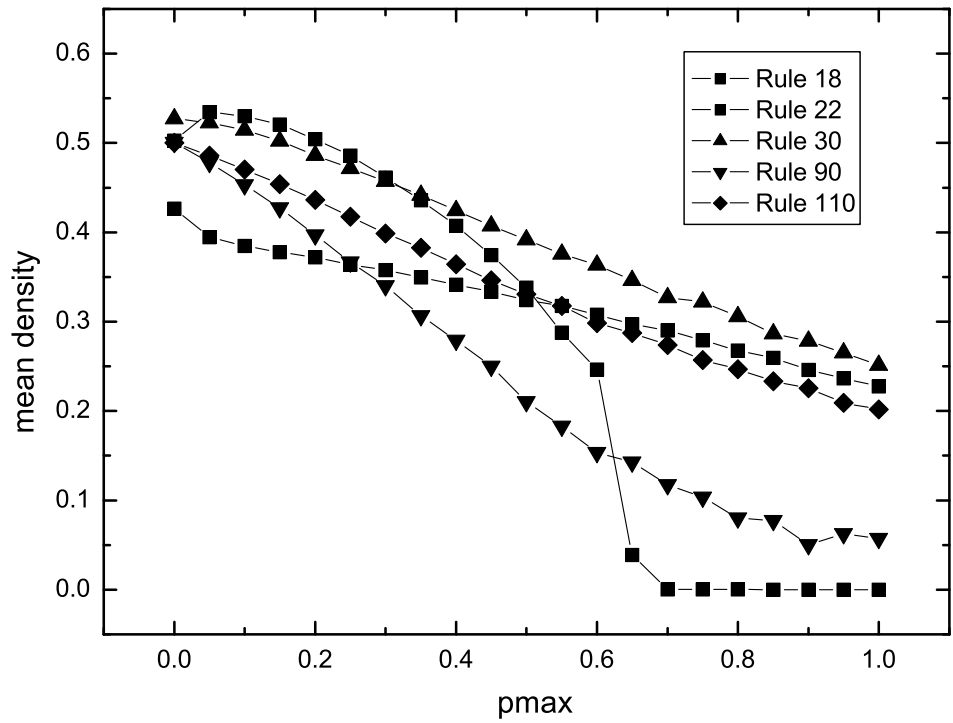}
\caption{Mean density $\rho$ vs. $pmax=\tilde p$ for different rules evolving 
under the second synchronization method. The existence
of a transition to a synchronized state can be clearly observed for rule $18$.}
\label{fig7}
\end{figure}

Now we introduce a new stochastic element in the application of the operator $\Gamma_p$. 
To differentiate from the previous case we call it $\tilde\Gamma_{\tilde p}$.  
The action of this operator consists in applying at each time the operator 
$\Gamma_p$, with $p$ chosen at random in the interval $(0,\tilde p)$. 
The evolution law of the automaton is in this case: 
\begin{equation}
	\sigma(t+1) = \tilde\Gamma_{\tilde p}[(\sigma^1(t), \sigma^2(t))] = 
	\tilde\Gamma_{\tilde p}[ (\sigma(t), \Phi[\sigma(t)]) ].
	\label{eq-auto1}
\end{equation}
The DA density between the present state and the precedent one, defined 
as $\delta(t)=\mid\sigma(t)-\sigma(t-1)\mid$, is plotted as a function of 
$\tilde p$ for different rules $\Phi$ in Fig. \ref{fig7}.
Only when the system becomes self-synchronized there will be a fall to zero in the DA density.
Let us observe again that the behavior commented in the first self-synchronization method is repeated
here.  Rule $18$ undergoes a phase transition for a critical value of $\tilde p$. 
For $\tilde p$ greater than such critical value, the method is able to find the stable structure
of the system. For the rest of the rules the freezing phase is not developed.
The dynamics generates patterns where the different marginally stable structures
randomly compete. This entails for the DA density to linearly decay with $\tilde p$
(see Fig. \ref{fig7}).

\subsection{A third self-synchronization method}

\begin{figure}
\includegraphics[angle=0, width=15cm]{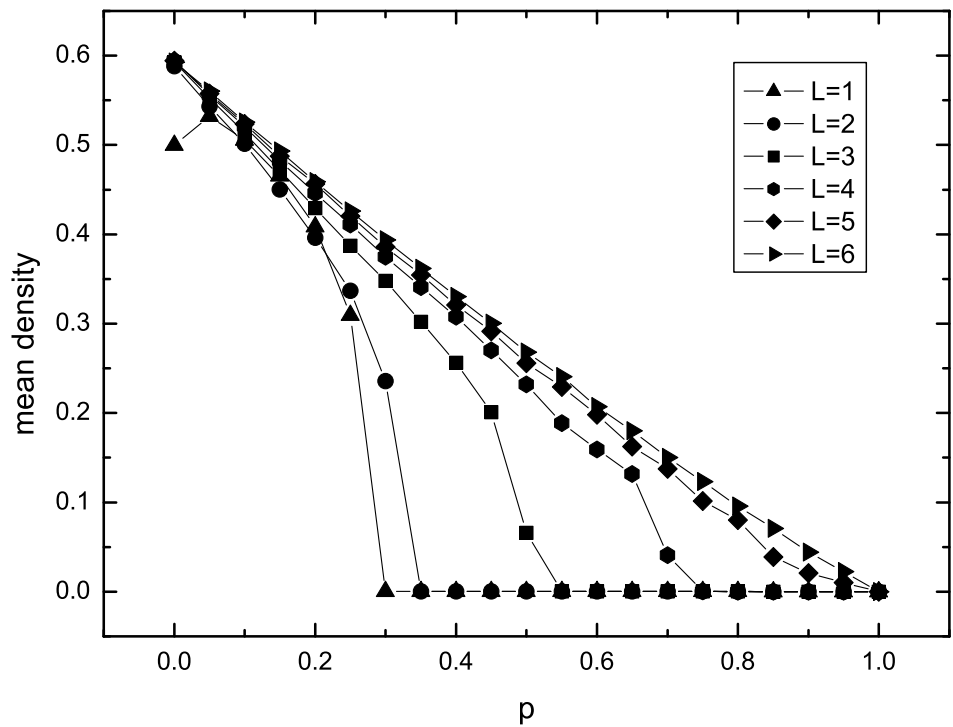}
\caption{Mean density $\rho$ vs. $p$ for rule $18$ evolving under
the third self-synchronization method.
The existence of a transition to a synchronized state can be observed despite of
the randomness in the election of neighbors within a range $L$, up to
$L=4$.}
\label{fig8}
\end{figure}

\begin{figure}
\includegraphics[angle=0, width=15cm]{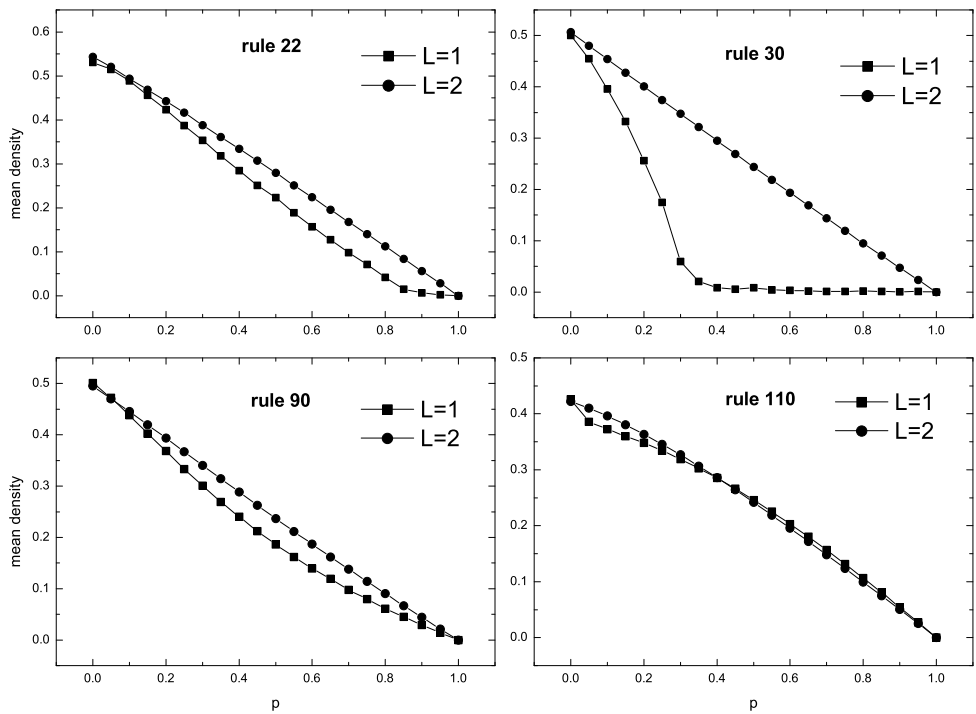}
\caption{Mean density $\rho$ vs. $p$ for different rules evolving 
under the third self-synchronization method. 
The density of the system decreases with $p$.}
\label{fig9}
\end{figure}

At last we introduce another type of stochastic element in the application of the rule $\Phi$. 
Given an integer number $L$, the surrounding of site $i$ at each time step is redefined. 
A site $i_l$ is randomly chosen among the $L$ neighbors of site $i$ to the left, $(i-L,\ldots,i-1)$.
Analogously, a site $i_r$ is randomly chosen among the $L$ neighbors of site $i$ to the right,
$(i+1,\ldots,i+L)$. The rule $\Phi$ is now applied on the site $i$ using the triplet $(i_l,i,i_r)$ 
instead of the usual nearest neighbors of the site. This new version of the rule is called $\Phi_L$,
being $\Phi_{L=1}=\Phi$.
Later the operator $\Gamma_p$ acts in identical way as in the first method.  
Therefore, the dynamical evolution law is:  
\begin{equation}
	\sigma(t+1) = \Gamma_p[(\sigma^1(t), \sigma^2(t))] = \Gamma_p[ (\sigma(t), \Phi_L[\sigma(t)]) ].
	\label{eq-auto2}
\end{equation}
The DA density as a function of $p$ is plotted in Fig. \ref{fig8} for the rule $18$
and in Fig. \ref{fig9} for other rules.
It can be observed again that the rule $18$ is a singular case that, 
even for different $L$, maintains the memory and continues to self-synchronize.
It means that the influence of the rule is even more important than the randomness 
in the election of the surrounding sites.
The system self-synchronizes and decays to the corresponding stable structure. 
Contrary, for the rest of the rules, 
the DA density decreases linearly with $p$ even for $L = 1$ as shown in Fig. \ref{fig9}. 
The systems oscillate randomly among their different marginally stable structures
as in the previous methods.

\section{Conclusions}
Inspired in stochastic synchronization methods for CA, different schemes 
for self-synchronization of a single automaton has been proposed and analyzed 
in this work.  Self-synchronization of a single automaton can be interpreted 
as a strategy for searching and controlling the stable structures of the system. 
In general, it has been found that a competition among all such structures is established,
and the system ends up oscillating randomly among them. 
However, rule $18$ is a singularity within this panorama,
since, even with random election of the neighbors sites, 
the automaton is able to reach the stable configuration.

\newpage

\end{document}